\documentclass[twocolumn,secnumarabic,amssymb, amsmath, nofootinbib,tightenlines,
nobibnotes, aps, prl,epsfig]{revtex4}
\usepackage{graphicx}
\usepackage{dcolumn}
\usepackage{bm}
\begin{document}
\preprint{APS/123-QED}
\title{Geometrical scaling in charm structure function ratios }

\author{G.R.Boroun}%
 \email{grboroun@gmail.com; boroun@razi.ac.ir }
\author{B.Rezaei }
\altaffiliation{brezaei@razi.ac.ir}
\affiliation{ Physics Department, Razi University, Kermanshah
67149, Iran}
\date{\today}
\begin{abstract}
By using a Laplace-transform technique, we solve the
next-to-leading-order master equation for charm production and
derive a compact formula for the
 ratio
 $R^{c}=\frac{F^{^{c\overline{c}}}_L}{F^{^{c\overline{c}}}_2}$,
 which is useful for extracting the
 charm structure function from the reduced charm cross section, in particular, at DESY HERA, at small $x$.
 Our results show that this ratio is  independent of $x$ at small $x$. In this method of determining the
 ratios, we apply geometrical scaling in charm production in
 deep inelastic scattering(DIS). Our analysis shows that the
renormalization scales have a sizable impact on the ratio $R^{c}$
at high $Q^{2}$. Our results for the ratio of the charm structure
functions are in
 good agreement with some phenomenological models.
\end{abstract}
\maketitle
\centerline{\textbf{1. Introduction}}

The $ep$ collider at HERA has played a crucial role in furthering
the understanding of the proton$^{,}$s structure. In the case of
pure photon exchange, the totally inclusive cross section of
deep-inelastic lepton-proton scattering (DIS) has the form
\begin{equation}
\frac{d^2\sigma}{dx dQ^2}={\frac{2\pi{\alpha}^2 Y_+}{Q^4
x}}.\sigma_r,
\end{equation}
where the reduced cross section can be defined by the structure
functions  $F_2 (x,Q^2)$ and $F_L (x,Q^2)$ as
\begin{equation}
\sigma_r\equiv F_2 (x,Q^2)-{\frac{y^2}{Y_+}}.F_L (x,Q^2),
\end{equation}
where $Y_+ =1+(1-y)^2$. The structure functions\quad $F_2$\quad
and\quad $F_L$\quad are related to the cross sections $\sigma_T$
and $\sigma_L$ for the interaction of transversely and
longitudinally polarized virtual photons with protons [1]. The
quark parton model (QPM) predicts $\sigma_L=0$, which leads to the
so-called Callan-Gross relation $F_L=0$, where it is further
broken by QCD corrections. Thus, in QCD, the longitudinal
structure function $F_{L}$ becomes non-negligible, and its
contribution should be properly taken into account when $F_{2}$ is
extracted from the measured cross section. However, the
contribution of the longitudinal structure function $F_L$ to the
cross section is sizeable only at large values of the inelasticity
y, and in most of the kinematic range, the relation
$\sigma_r\approx F_2$ holds to a very good approximation. The same
is true for the contributions $F^{^{c\overline{c}}}_2$ and
$F^{^{c\overline{c}}}_L$ to $F_{2}$
 and $F_{L}$ due to the charm quarks.\\
Therefore, precise measurements of the charm-inclusive scattering
cross section at the $ep$ collider are important for the
understanding of charmed meson  production. The charmed meson
production in deeply inelastic $ep$ scattering, in the
one-photon-exchange approximation, is via the reaction
\begin{equation}
e^{-}+p{\rightarrow}e^{-}+c\overline{c}+X.
\end{equation}
The reduced cross section is defined as
\begin{eqnarray}
\sigma^{c\overline{c}}_r&=&\frac{Q^4 x}{2\pi{\alpha}^2
Y_+}\frac{d^2\sigma^{c\overline{c}}}{dxdy}\nonumber\\
&&=F^{^{c\overline{c}}}_2
(x,Q^2,m^{2})-{\frac{y^2}{Y_+}}F^{^{c\overline{c}}}_L
(x,Q^2,m^{2})\nonumber\\
 &&=F^{^{c\overline{c}}}_2
(x,Q^2,m^{2})(1-{\frac{y^2}{Y_+}}R^{c}).
\end{eqnarray}
A measurement of the longitudinal charm structure function at
small $x$ at HERA is important because the
$F^{^{c\overline{c}}}_L$ contribution to the charm cross section
can be sizeable. At small values of $x$, $F^{^{c\overline{c}}}_L$
becomes non-negligible, and its contribution should be properly
taken into account when  $F^{^{c\overline{c}}}_2$ is extracted
from the measured charm cross section.\\
In perturbative QCD (pQCD) calculations, the production of heavy
quarks at HERA proceeds dominantly via  direct boson-gluon fusion
(BGF), where the photon interacts with a gluon from the proton
through the exchange of a heavy quark pair [2]. In recent years,
both the H1 and ZEUS collaborations have measured the charm
component $F^{^{c\overline{c}}}_2$ of the structure function at
small $x$ and have found it to be approximately ${\sim}30\%$ of
the total at HERA [3].\\
For the treatment of the charm component of the structure
function, there are basically two different prescriptions for
charm production in the literature. The first  is advocated in
[4], where the charm quark is treated as a heavy quark, and its
contribution is given by fixed-order perturbation theory. This
involves the computation of the boson-gluon fusion process. In the
other approach [5], the charm is treated similarly to a massless
quark, and its contribution is described by a parton density in a
hadron. Here, we consider the charm production via boson-gluon
fusion, where the charm is treated as a heavy quark and not a
parton. This scheme is usually called the fixed flavor number
scheme (FFNS). In this scheme, by definition, only light partons
(e.g. $u$, $d$, $s$ and $g$) are included in the initial state for
charm production and the number of parton flavors $n_{f}$ is kept
constant regardless of the energy scales involved. The boson-gluon
fusion gives the correct description of $F_{2}^{c}$ for $Q^{2} <
4m^{2}_{c}$ and should remain a reasonable approximation to
$F_{2}^{c}$ for $Q^{2}{\geq}4m^{2}_{c}$. However, the boson-gluon
fusion model will inevitably break down at larger $Q^{2}$ values
because the charm can no longer be treated as a non-partonic heavy
object, and begins to evolve in a similar manner to lighter
components of the quark sea. Therefore, our estimates in this
scheme should be considered with caution in the region of large
$Q^{2}$. A $c\overline{c}$ pair can be created by boson-gluon
fusion when the squared invariant mass of the hadronic final state
is in the region $W^{2}{\geq}4m^{2}_{c}$. Because $W^{2} =
Q^{2}(1-x) x +M^{2}_{N}$, where $M_{N}$ is the nucleon mass, charm
production can occur well below the $Q^{2}$ threshold, $Q^{2}
{\approx}4m^{2}_{c}$, at small $x$
[6].\\

In this paper, we investigate the NLO corrections to the
Callan-Gross ratio in heavy-quark leptoproduction, defined as
$R^{c}=\frac{F_{L}^{c}}{F_{2}^{c}}$, which is an observable that
is quantitatively well defined in perturbative quantum
chromodynamics (pQCD). Measurements of the quantity $R(x,Q^{2})$
in charm and bottom leptoproduction should provide a good test of
the conventional parton model based on pQCD. The leptoproduction
cross sections $\sigma^{{c\overline{c}}}_{k}(x,Q^{2})$ are related
to the structure functions $F^{{c\overline{c}}}_{k}(x,Q^{2})$, as
follows:
\begin{eqnarray}
F^{{c\overline{c}}}_{k}(x,Q^{2})=\frac{Q^{2}}{8\pi^{2}\alpha_{em}x}\sigma^{{c\overline{c}}}_{k}(x,Q^{2})
~~~~~~(k=T,L),
\end{eqnarray}
\begin{eqnarray}
F^{{c\overline{c}}}_{2}(x,Q^{2})=\frac{Q^{2}}{4\pi^{2}\alpha_{em}}\sigma^{{c\overline{c}}}_{2}(x,Q^{2}).
\end{eqnarray}
These hadron-level cross sections are related to the $\gamma^{*}g$
cross sections through the photon-gluon fusion mechanism [7-9],
and it is related to the virtual photon- proton cross section
$\sigma_{\gamma^{*}p}$. Data from deep inelastic scattering (DIS)
experiments at small $x$ exhibit an interesting property called
geometric scaling [10]. This means that the total ${\gamma^{*}p}$
cross section is not a function of the two variables $x$ and
$Q^{2}$ separately but rather a function of the combination
$Q^{2}/Q^{2}_{s}(x)$ only, where the "saturation scale" $Q_{s}$ is
defined such that saturation is expected to occur at $Q$ values
below $Q{_{s}}$ [11-12], as
\begin{equation}
\sigma_{\gamma^{*}p}(x,Q^{2})=\sigma_{\gamma^{*}p}(\tau),
\end{equation}
where $\tau=\frac{Q^{2}}{Q_{s}^{2}}$ is called a scaling variable
[13]. This scaling is a border between dense and dilute gluonic
systems. The saturation scale is customarily assumed to have a
power-like dependence on $x$, as  $Q^{2}_{s}(x)=Q^{2} _{0}
(x/x_{0})^{-\lambda}$, where $\lambda$ is a parameter that must be
determined from experimental data. In charm production,
geometrical scaling is expected to be violated due to the large
quark mass $m_{c}{\simeq}1.3GeV$. Therefore, the scaling variable
for charm production can be obtained by[13]
\begin{equation}
\tau_{c}=(1+\frac{4m_{c}^{2}}{Q^{2}})^{1+\lambda}\frac{Q^{2}}{Q_{0}^{2}}(\frac{x}{x_{0}})^{\lambda}.
\end{equation}
In this paper, we apply the quantity of geometrical scaling in the
ratio of the charm structure functions, $R^{c}$, for the NLO
analysis. Here, we extend the method proposed by the authors of
Refs.[14-17] by using a Laplace-transform technique for the
Dokshitzer-Gribov-Lipatov-Altarelli-Parisi (DGLAP) evolution
equations [18]. We derive a solution for the ratio of the charm
structure functions using the Laplace transform method for the
geometrical scaling at small $x$. However, our analysis shows that
the predictions for $R^{c}$ describe with good accuracy the small
$x$ predictions for NLO, and this analysis is directly independent
of the gluon distribution function. The structure of this article
is as follows. In Sect.2,  we briefly present the basic formalism
for the charm structure function at LO up to NLO. In Sect.3, we
use a Laplace-transform technique and predict  the ratio of the
charm structure functions at small $x$. Next, we use the
geometrical scaling in the ratio of the charm structure functions.
These results are discussed in
Sect.4.\\

\centerline{\textbf{2. Short theoretical input}}

The charm quark contributions
$F_{k}^{c}(x,Q^{2},m^{2}_{c})(k=2,L)$ to the proton structure
function at small $x$, where the gluon contribution is only
matter, are given by the following forms:
\begin{eqnarray}
F_{k}^{c}(x,Q^{2},m^{2}_{c})&=&2xe_{c}^{2}\frac{\alpha_{s}(\mu^{2})}{2\pi}\int_{ax}^{1}\frac{dy}{y^{2}}C_{g,k}^{c}
(\frac{x}{y},\zeta)\nonumber\\
&& {\times}G(y,\mu^{2}),
\end{eqnarray}
where $a=1+4\zeta(\zeta{\equiv}\frac{m_{c}^{2}}{Q^{2}})$,
$G(x,\mu^{2})$ is the gluon distribution function and $\mu$ is the
mass factorization scale, which has been set equal to the
renormalization scales $\mu^{2}=4m_{c}^{2}$ or
$\mu^{2}=4m_{c}^{2}+Q^{2}$. The $C^{c}_{g,k}$ are the charm
coefficient functions in the LO and NLO analysis as follows:
\begin{eqnarray}
C_{k,g}(z,\zeta)&{\rightarrow}&C^{0}_{k,g}(z,\zeta)+a_{s}(\mu^{2})[C_{k,g}^{1}(z,\zeta)\\\nonumber
&&+\overline{C}_{k,g}^{1}(z,\zeta)ln\frac{\mu^{2}}{m_{c}^{2}}],
\end{eqnarray}
where $a_{s}(\mu^{2})=\frac{\alpha_{s}(\mu^{2})}{4\pi}$. The
coefficient functions, in the LO analysis, can be determined [19]
as follows:
\begin{eqnarray}
C^{0}_{g,2}(z,\zeta)&=&\frac{1}{2}([z^{2}+(1-z)^{2}+4z\zeta(1-3z)-8{\zeta^{2}}z^{2}]\nonumber\\
&&{\times}ln\frac{1+\beta}{1-\beta}+{\beta}[-1+8z(1-z)\nonumber\\
&&-4z{\zeta}(1-z)]),
\end{eqnarray}
and
\begin{eqnarray}
C^{0}_{g,L}(z,\zeta)=-4z^{2}{\zeta}ln\frac{1+\beta}{1-\beta}+2{\beta}z(1-z),
\end{eqnarray}
where $\beta^{2}=1-\frac{4z\zeta}{1-z}$. In the NLO analysis, we
can use the compact form of these coefficients based on
 Refs.[8,20].\\

\subsection{3. Method }
We now derive the ratio of the  charm structure functions using a
Laplace-transform method that was used by BDHM [14-17]. The
coordinate transformation in this method is
\begin{eqnarray}
\upsilon &{\equiv}& \ln(1/x).
\end{eqnarray}
The functions $\hat{G}$, $\hat{C_{g,k}^{c}}$, and
$\hat{FF_{k}^{c}}$ in $\upsilon$-space are then defined by
\begin{eqnarray}
\hat{G}(\upsilon,Q^{2}) &{\equiv}& \hat{G}(e^{-\upsilon},Q^{2})
\end{eqnarray}
\begin{eqnarray}
\hat{C_{g,k}^{c}}(\upsilon) &{\equiv}&
\hat{C_{g,k}^{c}}(e^{-\upsilon})
\end{eqnarray}
\begin{eqnarray}
\hat{FF_{k}^{c}}(\upsilon,Q^{2}) &{\equiv}&
(2\frac{\alpha^{LO}_{s}(\mu^{2})}{2\pi}e^{2}_{c})^{-1}F_{k}^{c}.
\end{eqnarray}
 In this representation, Eq.(9) reduces to this form
\begin{eqnarray}
\hat{FF_{k}^{c}}(\frac{1}{a}\upsilon',Q^{2})=\int_{0}^{\upsilon'}\hat{G}(\omega,Q^{2})\\\nonumber
\frac{1}{a}e^{-(\upsilon'-\omega)}{C_{g,k}^{0}}(\frac{1}{a}e^{-(\upsilon'-\omega)})dw,
\end{eqnarray}
where
\begin{eqnarray}
\hat{H}_{k}(\upsilon')&{\equiv}&
\frac{1}{a}e^{-\upsilon'}{C_{g,k}^{0}}(\frac{1}{a}e^{-(\upsilon')}).
\end{eqnarray}
We introduce the notation that the Laplace transformation of a
function $\hat{H_{k}}(\upsilon')$ is given by $h_{k}(s)$, where
\begin{eqnarray}
h_{k}(s)&{\equiv}&
{\mathcal{L}}[\hat{H_{k}}(\upsilon');s]=\int_{0}^{\infty}\hat{H_{k}}(\upsilon')e^{-s\upsilon'}d\upsilon'.
\end{eqnarray}
We then reduce Eq.(17) to the form
\begin{eqnarray}
FF_{k}^{c}(s,Q^{2})&=&{\mathcal{L}}[\int_{0}^{\upsilon'}\hat{G}(\omega,Q^{2})\hat{H_{k}}(\upsilon'-\omega);s]\\\nonumber
&&=g(s,Q^2)h_{k}(s).
\end{eqnarray}
Let us introduce the ratio of the charm structure functions, which
is independent of the gluon distribution function in $s$ space, as
\begin{eqnarray}
\frac{F_{L}^{c}(s)}{F_{2}^{c}(s)}=\frac{h_{L}^{c}(s)}{h^{c}_{2}(s)}.
\end{eqnarray}
To determine a solution for this ratio, we use the following
property of the inverse Laplace transformation:
\begin{eqnarray}
{\mathcal{L}}^{-1}[F(s)G(s)]&=&\int_{0}^{t}f(t-\tau)g(\tau)d\tau\nonumber\\
&&=\int_{0}^{t}g(t-\tau)f(\tau)d\tau.
\end{eqnarray}
We then have
\begin{eqnarray}
{\mathcal{L}}^{-1}[{F_{L}^{c}(s)}{h^{c}_{2}(s)};\upsilon]={\mathcal{L}}^{-1}[{F_{2}^{c}(s)}{h^{c}_{L}(s)};\upsilon]
\end{eqnarray}
or
\begin{eqnarray}
\int_{0}^{\upsilon'}{F_{L}^{c}(w,Q^{2})}{\widehat{J}_{2}(\upsilon'-w)}dw\nonumber\\
=\int_{0}^{\upsilon'}{F_{2}^{c}(w,Q^{2})}{\widehat{J}_{L}(\upsilon'-w)}dw,
\end{eqnarray}
where
${\widehat{J}_{k}(\upsilon')}={\mathcal{L}}^{-1}[{h^{c}_{k}(s)};\upsilon'],
(k=2, L) $. Based on the ratio of the charm structure functions,
$R^{c}(x,Q^{2})=\frac{F_{L}^{c}}{F_{2}^{c}}$, one obtains
\begin{eqnarray}
\int_{0}^{\upsilon'}R^{c}(w,Q^{2}){F_{2}^{c}(w,Q^{2})}{\widehat{J}_{2}(\upsilon'-w)}dw\nonumber\\
=\int_{0}^{\upsilon'}{F_{2}^{c}(w,Q^{2})}{\widehat{J}_{L}(\upsilon'-w)}dw.
\end{eqnarray}
The ratio $R^{c}$ was previously studied in the framework of the
$k_{t}$-factorization approach and determined to be approximately
$x$ independent in the small $x$ region. Indeed, it has been found
that at small $x$ and small $Q^2$ (where approaches based on
perturbative QCD, and on $k_t$ factorization give similar
predictions [8]) ratio $R^c$ is quite flat. On the other hand at
small $x$ and high $Q^2$, $R^c$ rises in the framework of
perturbative QCD. This can be due to the small-$x$ re-summation,
which is important at high
$Q^{2}$ [7-9].\\
Assuming the small-$x$ ($x{\leq}0.01$) behavior of the ratio to be
$R^{c}(x{\rightarrow}0,Q^{2}){\equiv}R^{c}(Q^{2})$, we consider
compact formulae for this ratio of the charm structure functions
in this region of $x$ values. Based on the assumption, we
transformed to $\upsilon$-space, where it lies in the interval
$\eta<\upsilon<\upsilon'$ (where $\eta$ satisfies the boundary
condition for $x{\leq}0.01$). Therefore,
\begin{eqnarray}
R^{c}(Q^{2})=\frac{\int_{\eta}^{\upsilon'}{F_{2}^{c}(w,Q^{2})}{\widehat{J}_{L}(\upsilon'-w)}dw}
{\int_{\eta}^{\upsilon'}{F_{2}^{c}(w,Q^{2})}{\widehat{J}_{2}(\upsilon'-w)}dw}.
\end{eqnarray}
The integrations in Eq. (26) are then taken to run from $y=ax$ to
$y=0.01$ rather than from $y=ax$ to $y=1$. Finally, we have the
ratio of the charm structure functions in $x$-space as follows:
\begin{eqnarray}
R^{c}(Q^{2})=\frac{\int_{ax}^{0.01}{\frac{x}{y}F_{2}^{c}(y,Q^{2})}C_{g,L}^{c}
(\frac{x}{y},\zeta)\frac{dy}{y}}
{\int_{ax}^{0.01}{\frac{x}{y}F_{2}^{c}(y,Q^{2})}C_{g,2}^{c}
(\frac{x}{y},\zeta)\frac{dy}{y}}.
\end{eqnarray}
This result  is in the framework of the $k_{t}$- factorization
approach at small $x$. Based on Eq. (6), the charm structure
function $F_{2}^{c\overline{c}}$ is related to the photon-proton
cross section for charm production by the simple relation
$\sigma^{{c\overline{c}}}_{2}(x,Q^{2})=4\pi^{2}\alpha_{em}F^{{c\overline{c}}}_{2}(x,Q^{2})/{Q^{2}}$.
We know that the geometrical scaling hypothesis for charm
production means that
\begin{equation}
\sigma^{{c\overline{c}}}_{2}(x,Q^{2})=\frac{1}{Q_{0}^{2}}f{(\tau_{c})},
\end{equation}
where $f$ is a universal, dimensionless function of the scaling
variable [13]. The constant $Q_{0}^{2}$ sets the dimension, and it
can be extracted from the data. To determine this function, one
should know
 the cross section $\sigma_{dp}(u)$  describing the
interaction of the $q\overline{q}$ color dipole with the proton
and doing the respective integration for the $\sigma_{\gamma^{*}
p}$. Because $\sigma_{dp}$ is not calculable, some models [13],
such as $1-\exp(-u^2)$, are used, where those values in this model
are compared  with the experimental data. In contrast, there exist
some parameterizations of $F^{c\overline{c}}_{2}$ whereby the
function $f$ can be calculated by dividing
the parameterization $F^{c\overline{c}}_{2}$ by $Q^{2}$ [12].\\
 Therefore, the charm structure function
is scaled by  the geometrical scale as
\begin{eqnarray}
F^{{c\overline{c}}}_{2}(x,Q^{2})=\frac{1}{4\pi^{2}\alpha_{em}}\frac{Q^{2}}{Q_{0}^{2}}f{(\tau_{c})}.
\end{eqnarray}
Finally, the ratio of the charm structure functions can be
obtained by the following form:
\begin{eqnarray}
R^{c}(Q^{2})=\frac{\int_{ax}^{0.01}\frac{x}{y}f(\tau_{c})C_{g,L}^{c}
(\frac{x}{y},\zeta)\frac{dy}{y}}
{\int_{ax}^{0.01}\frac{x}{y}f(\tau_{c})C_{g,2}^{c}
(\frac{x}{y},\zeta)\frac{dy}{y}}.
\end{eqnarray}

\subsection{4. Results and Conclusion }
In this section, we intend to use Eq. (30) to extract the ratio of
the charm structure functions at small $x$ at NLO. The parameters
$(\sigma_{0}, x_{0})$ were obtained from a fit to the HERA data
[21-22]. We take $\lambda_{c}=0.558{\pm}0.038$, which is an
optimal value of the parameter $\lambda$ in charm production for
$m_{c}=1.3~GeV$ (as accompanied with statistical error )[13]. We
set the running coupling constant to $\Lambda=0.224~ GeV$, and the
theoretical uncertainties in our result are based on the
renormalization scales $\mu^{2}=4m^{2}_{c}$ and
$\mu^{2}=4m^{2}_{c}+Q^{2}$.\\
In Fig. 1, we present our results for the ratio
$R^{c}=F_{L}^{c}/F_{2}^{c}$ in charm leptoproduction at NLO. We
observe that this ratio is independent of $x$ for $x~{\leq}~0.01$
in a wide range of $Q^{2}$. We see that this value is
approximately between $0.14$ and $0.18$ in a region of $Q^2$, and
this prediction for $R^{c}$ is nearly equal to the results in
Refs.[7-9]. While the NLO result for $R^{c}$ is independent of $x$
at small $x$, we can make use of Eq.(26) and impose the geometric
scaling condition in charm production based on Eq.(30). The
solution of this equation is general, and the only observation
that we can make is that it should generate $\tau_{c}$ behavior
with respect to $\lambda_{c}$.\\
In Fig. 2, we present the ratio $R^{c}$ as a function of $Q^{2}$
at $x{\leq}0.01$ from Eq.(30) with respect to $\mu^{2}=4m^{2}_{c}$
and $\mu^{2}=4m^{2}_{c}+Q^{2}$ for the  geometrical scaling in
charm production. We can see that the behavior of this ratio is in
good agreement  with the prediction from Refs.[7-9] because both
have a maximum value between $Q^{2}=10$ and $100 ~GeV^{2}$ and
then fall as $Q^{2}$ increases. These results are in agreement
with the $\mathrm{k_{t}}$- factorization approach [23] only at
small $Q^{2}$, that is, it continues to rise as $Q^{2}$ increases.
At high $Q^{2}$ values, our results are dependent on the
renormalization scales. In Fig. 3, we compare our results at the
renormalization scale $\mu^{2}=4m^{2}_{c}+Q^{2}$ with the results
in Ref.[7] (N.Ya.Ivanov and B.A.Kinehl, Eur.Phys.J.C{\bf59},
647(2009)) for the ratio $R^{c}$ because the authors derived an
analytical small-$x$ formula with arbitrary values of $\delta$ in
terms of the Gauss hypergeometric function. The value of $\delta$
is based on the gluon behavior. The prediction has the following
form [7]:
\begin{widetext}
\begin{eqnarray}
R^{(\delta)}(Q^{2})=4\frac{\frac{2+\delta}{3+\delta}\Phi(1+\delta,\frac{1}{1+4\zeta})-(1+4\zeta)\Phi(2+\delta,\frac{1}{1+4\zeta})}
{[1+\frac{\delta(1-\delta^{2})}{(2+\delta)(3+\delta)}]\Phi(\delta,\frac{1}{1+4\zeta})-(1+4\zeta)(4-\delta-\frac{10}{3+\delta})\Phi(1+\delta,\frac{1}{1+4\zeta})},
\end{eqnarray}
\end{widetext}
where the function $\Phi(r,z)$ is defined as
\begin{eqnarray}
\Phi(r,z)=\frac{z^{1+r}}{1+r}\frac{\Gamma(1/2)\Gamma(1+r)}{\Gamma(3/2+r)}
{_{2}F_{1}}(\frac{1}{2},1+r;\frac{3}{2}+r;z),\nonumber\\
\end{eqnarray}
and the hypergeometric function ${_{2}F_{1}}(a,b;c;z)$ has the
following series expansion:
\begin{eqnarray}
{_{2}F_{1}}(a,b;c;z)=\frac{\Gamma(c)}{\Gamma(a)\Gamma(b)}\sum_{n=0}^{\infty}\frac{\Gamma(a+n)\Gamma(b+n)}{\Gamma(c+n)}\frac{z^{n}}{n!}.
\end{eqnarray}
For $x<0.01$, our results are compatible with the results of
Ref.[7] because $\delta{\rightarrow}~0$, and when $x{=}0.01$, our
results are compatible with the case where
$\delta{\rightarrow}~\frac{1}{2}$. We observe that our results are
general and converge to other results over the entire $Q^{2}$
range [7-9,23-25]. In Figs.2 and 3, the $Q^{2}$ dependence of the
ratio $R^{c}$ is investigated. We conclude that the hadron-level
predictions for $R^{c}(x\rightarrow 0,Q^{2})$ are stable not only
under the NLO corrections without having knowledge about the gluon
distribution behavior but also under the Laplace transform
technique in the limit of the geometrical scaling in charm
production. In addition, the ratio $R^{c}$ could be a probe of the
charm density in the proton at
$x{\leq}~0.01$ and at high $Q^{2} >> m_{c}^{2}$.\\
In conclusion, we have tried to determine the ratio of the charm
structure functions using a Laplace transform technique for the
geometrical scaling in charm production in deep inelastic
scattering for small $x$. Our result is model independent for the
gluon distribution behavior, and it is dependent on the running
coupling constant for the NLO analysis and the renormalization
scales. Our results show that the ratio of the charm structure
functions is valid up to $x{\leq}0.01$, and this is well domain of
geometrical scaling within the saturation models. We suggest that
this method is useful for the extraction of $F_{2}^{c}$ from the
corresponding reduced cross section because it is
insensitive to the gluon behavior in the QCD input parameters.\\
\subsection{Acknowledgment}
The authors would like to thank P.Ha and T.Stebel for their
helpful comments and
useful suggestions.\\
\section{References}

1. D.H.Perkins,{Introduction to High Energy Physics},(University
of Oxford,Oxford, England,1982),ADDISON-WESLEY PUBLISHING ;
 Francis Halzen and Alan D.Martin,{Quarks and Leptons},(JOHN WILEY \&
 SONS,1984).\\
2. K.Lipka, Pos(EPS-HEP),313(2009).\\
3. C. Adloff et al. [H1 Collaboration], Z. Phys. C\textbf{72},
593(1996); J. Breitweg et al. [ZEUS Collaboration], Phys. Lett.
B\textbf{407}, 402(1997); C. Adloff et al. [H1 Collaboration],
Phys. Lett. B\textbf{528}, 199(2002); S. Aid et. al., [H1
Collaboration], Z. Phys. C\textbf{72}, 539(1996); J. Breitweg et.
al., [ZEUS Collaboration], Eur. Phys. J. C\textbf{12}, 35(2000);
S. Chekanov et. al., [ZEUS Collaboration], Phys. Rev.
D\textbf{69}, 012004(2004); Aktas et al. [H1 Collaboration], Eur.
Phys.J. C\textbf{45}, 23(2006); F.D. Aaron et al. [H1
Collaboration],Eur.Phys.J.C\textbf{65}, 89(2010).\\
4. M. Gluck, E. Reya, M. Stratmann. Nucl. Phys.B\textbf{422}, 37(1994); J.Blumlein, et al., Nucl. Phys.B\textbf{755}, 272(2006).\\
5. M. A. G. Aivazis et al. Phys.Rev.D\textbf{50}, 3102(1994).\\
6. A.L.Ayala Filho, M.B. Gay Ducati and Victor P.Goncalves, Phys.Rev.D\textbf{59}, 054010(1999).\\
7.  N.Ya.Ivanov, Nucl.Phys.B\textbf{814} , 142 (2009);
N.Ya.Ivanov and B.A.Kinehl, Eur.Phys.J.C{\bf59}, 647(2009); arXiv:hep-ph/1212.3785, (2012); arXiv:hep-ph/1212.3783, (2012).\\
8. A.~Y.~Illarionov,B.~A.~Kniehl and A.~V.~Kotikov, Phys.Lett. B {\bf 663}, 66 (2008); A.V.Kotikov, et.al., Eur.Phys.J.C{\bf26}, 51(2002).\\
9. I.P.Ivanov and N.Nikolaev,Phys.Rev.D{\bf65},054004(2002); \\
10. A. M. Stasto, K. Golec-Biernat, and J. Kwiecinski Phys.Rev.
Lett.{\textbf86}, 596(2001); K. Golec-Biernat, J.Phys.G{\textbf28}, 1057(2002); K. Golec-Biernat, Acta Phys.Pol.B{\textbf35}, 3103(2004). \\
11. E.Avsar and G.Gustafson, JHEP{\textbf0704}, 067(2007).\\
12. G.Beuf, C.Royon and D.Salek, arXiv:0810.5082(2008); C.Royon
and
R.Peschanski; PoS DIS 2010, 282(2010)(arXiv:hep-ph/1008.0261).\\
13. T.Stebel, arXiv:hep-ph/1305.2583(2013); M.Praszalowicz and
T.Stebel,
JHEP {\textbf03}, 090(2013); T. Stebel, Master Thesis, arXiv:hep-ph/1210.1567(2012); M.Praszalowicz,  arXiv:hep-ph/1304.1867(2013)\\
14.M.M.Block, L.Durand, D.W.McKay, Phys.Rev.D\textbf{77},
094003(2008).\\
15.M.M.Block, L.Durand, D.W.McKay, Phys.Rev.D\textbf{79},
014031(2009).\\
16.M.M.Block, Eur.Phys.J.C\textbf{65},
1(2010).\\
17.M.M.Block, L.Durand, Phuoc Ha, D.W.McKay,
Phys.Rev.D\textbf{83},
054009(2011).\\
18. Yu. L.Dokshitzer, Sov.Phys.JETPG {\bf6}, 641(1977 );
G.Altarelli and
G.Parisi, Nucl.Phys.B{\bf126}, 298(1997 ); V.N.Gribov and L.N.Lipatov, Sov.J.Nucl.Phys.{\bf28}, 822(1978).\\
19. M.Gluk, E.Reya and A.Vogt, Z.Phys.C\textbf{67}, 433(1995); Eur.Phys.J.C\textbf{5}, 461(1998).\\
20. S. Catani, M. Ciafaloni and F. Hautmann, Preprint
CERN-Th.6398/92, in Proceeding of the Workshop on Physics at HERA
(Hamburg, 1991), Vol. 2., p. 690; S. Catani and F. Hautmann, Nucl.
Phys. B \textbf{427}, 475(1994); S. Riemersma, J. Smith and W. L.
van Neerven, Phys. Lett. B \textbf{347}, 143(1995).\\
21. K. Golec-Biernat and M.Wusthoff, Phys.Rev.D{\textbf59}, 014017(1999). \\
22. M.B.Gay Ducati, M.N.Machado and M.V.T.Machado, Brazilian
J.Phys,{\textbf38}, 487(2008); V.P.Goncalves and M.V.T.Machado,
Phys.Rev.Lett.{\textbf91}, 202002(2003).\\
23. A.~V.~Kotikov, A.~V.~Lipatov, G.~Parente and N.~P.~Zotov Eur.\
Phys.\ J.\  C {\bf 26}, 51 (2002).\\
24. Carlo Ewerz, et.al., arXiv:hep-ph/1201.6296(2012).\\
25. G.R.Boroun, B.Rezaei, JETP,Vol.115, No.7, PP.427 (2012);
Nucl.Phys.B{\bf857}, 143(2012); Eur.Phys.J.C{\bf72}, 2221 (2012);
EPL{\bf100},41001(2012).\\

\begin{figure}
\includegraphics[width=1\textwidth]{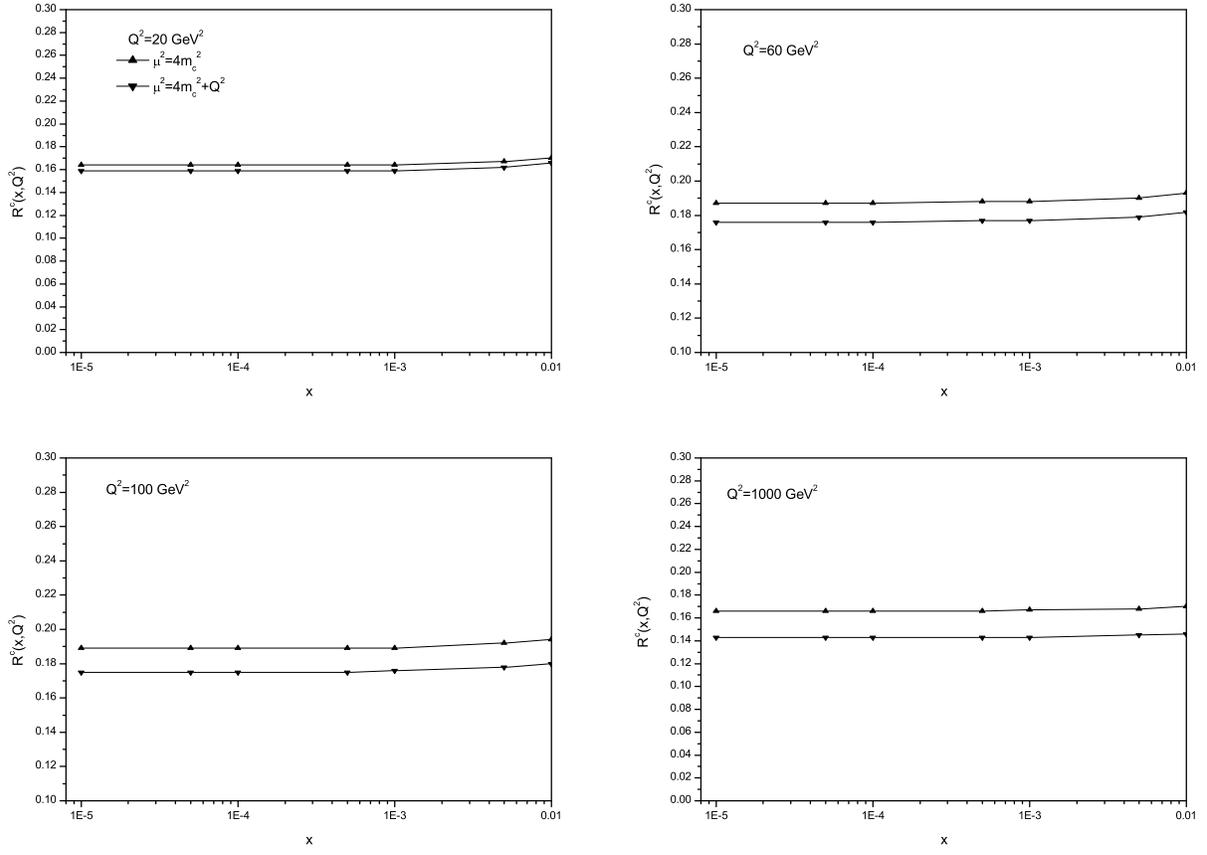}
\caption{The ratio $R^{c}=F_{L}^{c}/F_{2}^{c}$ as a function of
$x$ for different values of $Q^{2}$ in NLO analysis.} \label{Fig1}
\end{figure}
\begin{figure}
\includegraphics[width=1\textwidth]{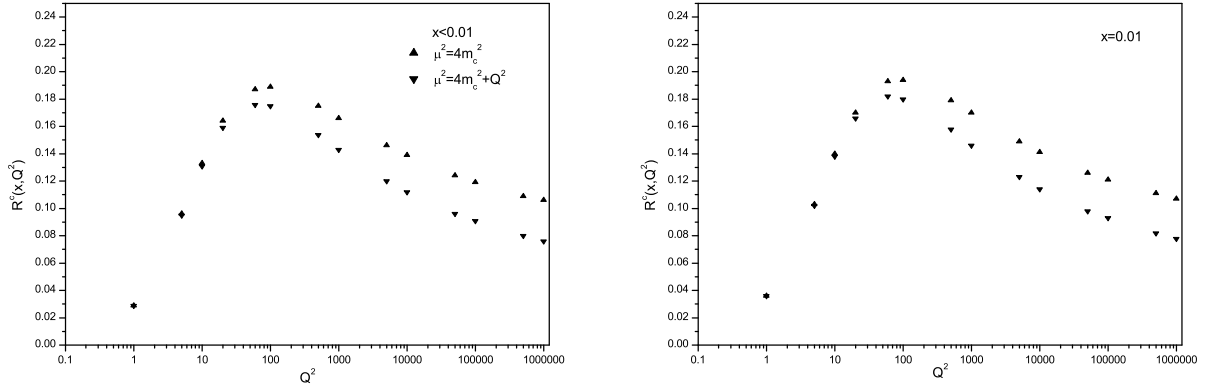}
\caption{The ratio $R^{c}=F_{L}^{c}/F_{2}^{c}$ as a function of
 $Q^{2}$ for $x{\leq}0.01$. Plotted are the NLO predictions at the renormalization scales $\mu^{2}=4m^{2}_{c}$ and $\mu^{2}=4m^{2}_{c}+Q^{2}$. } \label{Fig2}
\end{figure}
\begin{figure}
\includegraphics[width=1\textwidth]{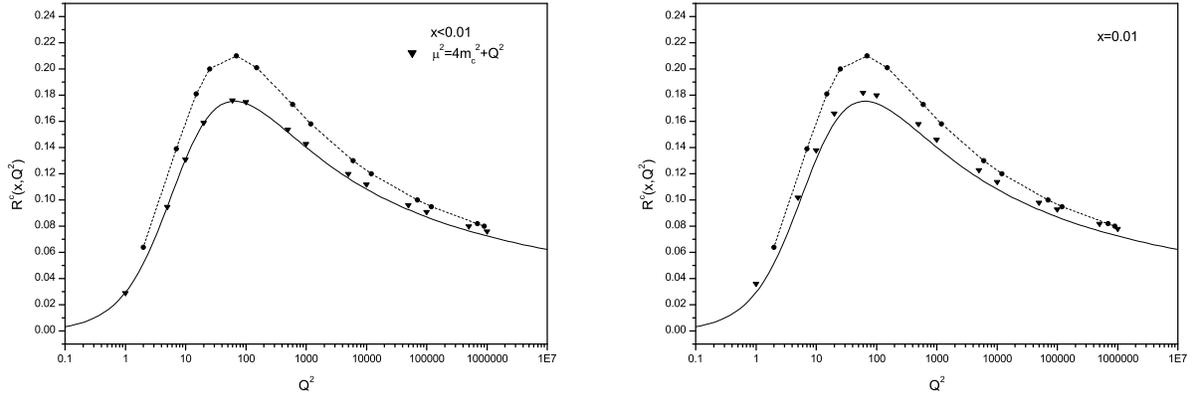}
\caption{The same result as in Fig. 2 at the renormalization scale
 $\mu^{2}=4m^{2}_{c}+Q^{2}$ compared to the Gauss hypergeometric function (Eq.31) [7]. The solid curve is the asymptotic ratio at
 $\delta=0$,
 and the dot-point curve is at $\delta=0.5$ [7].  }
\label{Fig3}
\end{figure}

\end{document}